\documentclass[aps, pra,twocolumn, amsmath,amssymb,superscriptaddress, 10pt
]{revtex4-2}

\usepackage{graphicx} 

\usepackage{mathptmx} 
\usepackage{xcolor}   
\usepackage{physics}  
\usepackage{dcolumn}
\usepackage{tabularx}

\usepackage{hyperref} 
\hypersetup{
    colorlinks=true,
    linkcolor=blue,
    citecolor=blue,
    filecolor=blue,
    urlcolor=blue
}

\newcommand{\UMD}{University of Maryland, College Park, Maryland, USA}
\newcommand{\SSD}{Sensor Science Division, National Institute of Standards and Technology, Gaithersburg, Maryland 20899, USA}
\newcommand{\CTL}{Communications Technology Laboratory, National Institute of Standards and Technology, Boulder, Colorado 80305, USA}

\begin{document}


\author{David S. La Mantia}
\affiliation{\SSD}

\author{Mingxin Lei}
\affiliation{\UMD}

\author{Nikunjkumar Prajapati}
\affiliation{\CTL}


\author{Noah~Schlossberger}
\affiliation{\CTL}

\author{Matthew T. Simons}
\affiliation{\CTL}

\author{Christopher L. Holloway}
\affiliation{\CTL}

\author{Julia  Scherschligt}
\affiliation{\SSD}

\author{Stephen P. Eckel}
\affiliation{\SSD}

\author{Eric B. Norrgard}
\email[Contact author: ]{eric.norrgard@nist.gov}
\affiliation{\SSD}
 
\title{Compact Blackbody Radiation Atomic Sensor:\break Measuring Temperature using Optically Excited Atoms in Vapor Cells}
\begin{abstract}
We demonstrate a blackbody radiation thermometer based on optically excited rubidium atoms in a vapor cell.  The temperature measurement is fast, with statistical uncertainty as low as 0.1~\% in one second. 
We resolve temperature with a precision of 0.04\,\% in the range 308\,K to 344\,K when averaging for several seconds.    Additionally, we describe an extension to this measurement scheme where the device operates as a self-calibrated, or primary, thermometer.  We make progress toward realizing a primary thermometer by demonstrating a temperature-dependent self-consistent  calibration scheme, with temperature accuracy of order 1~\% limited by the uncertainty in atomic transition dipole matrix elements.   

\end{abstract}

\maketitle
Thermocouples, resistance thermometers, liquid-in-glass thermometers, infrared non-contact thermometers, and many other thermometer designs are actually ``proxy thermometers'': they measure a temperature-dependent parameter rather than temperature as defined by the SI.
Such proxy thermometers require ongoing periodic calibration to ensure accuracy.
Calibrations can be costly and time-consuming, so creating a calibration-free device that directly measures temperature would be of great benefit to the end-user.
Here, we propose a self-calibrating, or ``primary'', thermometer based on the intensity ratios of fluorescence measured in a laser-excited Rb atom and experimentally demonstrate essential components thereof.

Fluorescence intensity ratios of luminescent rare-earth-doped and transition metal-doped materials can sense temperature changes~\cite{Wang2021,HAN2019,Jia2020,Li2021}.
However, the response of these materials is not calculable from first principles at sufficient accuracy to provide a temperature reading without calibration.
Isolated atoms and small molecules are comparatively simple and well-characterized quantum systems which may be used as absolute probes of radiometric~\cite{Norrgard2021, Schlossberger2024} and thermodynamic~\cite{Gotti2021} temperature.  
For example, recent studies of microwave sensing using spontaneous \cite{Prajapati2024} and stimulated (six-wave mixing) \cite{Borowka2024} decay fluorescence detection in Rydberg atoms have found that the noise floor is consistent with residual excitation due to room-temperature blackbody microwave radiation.

In this letter, we experimentally realize a thermometer using fluorescence intensity ratios of optically-excited Rb atoms in a vapor cell.
Our Compact Blackbody Radiation Atomic Sensor \mbox{(CoBRAS)} resolves radiometeric temperature  to within 0.04\,\%  by comparing an observed fluorescence ratio to a simple rate equation model.
As a step toward realizing a primary thermometer, we also demonstrate calibration of the CoBRAS by subsequently optically exciting to an additional atomic state.
This technique tests the accuracy and self-consistency of our rate equation model, which we find to be limited by the theoretical accuracy of the transition dipole matrix elements (TDMEs).
We estimate the magnitude of other potential systematic errors and highlight potential future directions of this temperature measurement technique, including primary self-calibration.

\begin{figure}
    \centering
    \includegraphics[width = .9\columnwidth]{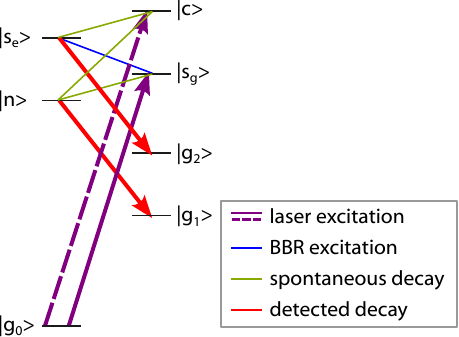}
    \caption{Schematic level diagram of fluorescence thermometry. The $\ket{g_0}\rightarrow\ket{s_g}$ transition is driven by a laser, while $\ket{n}\rightarrow\ket{g_1}$ fluorescence and  $\ket{s_e}\rightarrow\ket{g_2}$ fluorescence are monitored.  BBR excites the $\ket{s_g} \rightarrow\ket{s_e}$ transition with a temperature dependent rate.  Other spontaneous decays rates are essentially temperature-independent.  Detectors may be calibrated without knowing the radiometric temperature by driving the $\ket{g_0}\rightarrow\ket{c}$ transition while monitoring the same fluorescence transitions. }
    \label{fig:concept}
\end{figure}

Our method for measuring blackbody radiation (BBR) uses a driven, multi-level quantum system in steady state and is illustrated in Fig.\,\ref{fig:concept}.
The system is driven from the ground or low-lying metastable state $\ket{g_0}$ to the ground sensing state $\ket{s_g}$. 
The system measures BBR at the transition wavelength $\lambda_{s_e, s_g}$ between the ground sensing state $\ket{s_g}$ and the excited sensing state $\ket{s_e}$ due to BBR-induced excitation.
Excitation of the $\ket{s_g}\rightarrow\ket{s_e}$ transition is detected by subsequent fluorescence from $\ket{s_e}$ decaying to a lower energy state $\ket{g_2}$.
Absent other excitation mechanisms, the rate of fluorescence at wavelength $\lambda_{s_e, g_2}$ is due entirely to the presence of BBR at wavelength $\lambda_{s_e, s_g}$ and will depend on the radiometric temperature.
We refer to this as ``signal fluorescence''.

Simultaneously, the ground sensing state $\ket{s_g}$ may decay to some normalization state $\ket{n}$.  The $\ket{s_g}\rightarrow \ket{n}$ transition rate is temperature-independent to the extent that the spontaneous decay rate  exceeds the BBR-stimulated and collision-induced decay rates for this transition.
This condition is satisfied whenever $\ket{n}$ lies below $\ket{s_g}$ in energy and $hc/\lambda_{n,s_g}\gtrsim k_B T$, where $h$ is the Planck constant and $c$ is the speed of light.
 This temperature-independent transition is detected by subsequent fluorescence from $\ket{n}$ decaying to some lower energy state $\ket{g_1}$.  
 We refer to this as ``normalization fluorescence''.
We note that while it is conceptually helpful to consider the normalization transition rate as temperature-independent, this condition is not necessary for fluorescence intensity ratio thermometry, and the small temperature dependence is accounted for in this work.

 Subsequent to detecting signal and normalization fluorescence while driving the $\ket{g_0} \rightarrow \ket{s_g}$ transition, the procedure is repeated when driving the calibration transition  $\ket{g_0} \rightarrow \ket{c}$.  State $\ket{c}$ is chosen to lie above $\ket{s_e}$ and $\ket{n}$ in energy so that the $\ket{c}\rightarrow \ket{s_e}$ and $\ket{c} \rightarrow \ket{n}$ transition rates are essentially temperature independent.  We refer to this as the ``self-calibration criterion''.  When driving the calibration transition, the signal and normalization fluorescence is  used to calibrate the fluorescence detector using the intrinsically stable properties of the multi-level quantum system. Crucially, to the extent that the $\ket{c}\rightarrow \ket{s_e}$ and $\ket{c} \rightarrow \ket{n}$ transition rates are temperature-independent, this calibration does not require knowledge of the radiometric temperature $T$ at which the detector was calibrated. 
 %

 When optically driving an ensemble of atoms on the transition from the ground state $\ket{g_0}$ to excited state \mbox{$\ket{k}= \{ \ket{s_g}, \ket{c} \}$}, the rate of detecting fluorescence photons with transition wavelength $\lambda_{ij}$ is
 \begin{equation}
     S_{\lambda_{ij}}^{(k)}  = N \Gamma_{ij} \eta_{\lambda_{ij}} p_i^{(k)}(T),
 \end{equation}
where $N$ is the total number of atoms, $\Gamma_{ij}$ is the decay rate from $\ket{i}$ to $\ket{j}$, and $\eta_{\lambda_{ij}}$ is the total detection efficiency at wavelength $\lambda_{ij}$. The fractional atomic population in state $\ket{i}$  is written as  $ p_i^{(k)}(T)$  to emphasize that it may explicitly depend upon the radiometric temperature  $T$ and the optically excited state $\ket{k}$.  
The radiometric temperature may be determined from the ratio of the simultaneously-measured signal and normalization fluorescence when exciting to the ground sensing state $\ket{s_g}$:
\begin{equation}\label{eq:ratio}
\begin{split}
      r^{(s_g)}_{\lambda_{s_e,g_2}, \lambda_{n,g_1} } (T)&= S_{\lambda_{s_e,g_2}} ^{(s_g)} / S_{\lambda_{n,g_1}}^{(s_g)} \\
      & = \frac{\eta_{\lambda_{s_e,g_2}}}{ \eta_{\lambda_{n,g_1} } } 
      \frac{\Gamma_{s_e,g_2} p^{(s_g)}_{s_e}(T) }  {\Gamma_{n,g_1} p^{(s_g)}_{n}(T) }.
\end{split}
\end{equation}
In \eqref{eq:ratio}, the fluorescence ratio $ r^{(s_g)}_{\lambda_{s_e,g_2}, \lambda_{n,g_1} }$ has been written as the product of a detector-dependent factor, $\eta_{\lambda_{s_e,g_2}}/ \eta_{\lambda_{n,g_1} }$, and an atom- and temperature-dependent factor $\Gamma_{s_e,g_2} p^{(s_g)}_{s_e}(T) /[\Gamma_{n,g_1} p^{(s_g)}_{n}(T)]$.
Assuming the atom-dependent term is well known, the detector-dependent term must be determined in order to properly calibrate the temperature response of the CoBRAS.
Such a calibration can be done in at least three ways.
First, one could determine the detector-dependent factor to an accuracy of 0.005\,\% using radiometric transfer standards~\cite{Houston2022,Norrgard2023}.
Second, one could infer the ratio of total detection efficiencies from the observed $r^{(s_g)}_{\lambda_{s_e,g_2}, \lambda_{n,g_1} }$ at one or more known temperatures.

Finally, if the atom-dependent term is well-known for both $\ket{s_g}$ and $\ket{c}$, the detector-dependent factor may be determined by driving the calibration transition, such that
\begin{equation}\label{eq:ratio of ratios}
\begin{split}
     r^{(s_g)}_{\lambda_{s_e,g_2}, \lambda_{n,g_1} } (T)
      &= r^{(c)}_{\lambda_{s_e,g_2}, \lambda_{n,g_1} } (T)  \frac {\Gamma_{\lambda_{n,g_1}} p^{(c)}_{n}(T) } {\Gamma_{s_e,g_2} p^{(c)}_{s_e}(T) } 
      \frac{\Gamma_{s_e,g_2} p^{(s_g)}_{s_e}(T) }  {\Gamma_{\lambda_{n,g_1}} p^{(s_g)}_{n}(T) }\\
    &=r^{(c)}_{\lambda_{s_e,g_2}, \lambda_{n,g_1} }   \frac {p^{(c)}_{n} } { p^{(c)}_{s_e} } 
      \frac{ p^{(s_g)}_{s_e}(T) }  { p^{(s_g)}_{n} }.
\end{split}\end{equation}
In the second line of \eqref{eq:ratio of ratios}, it is assumed that only $ p^{(s_g)}_{s_e}$ strongly depends on radiometric temperature.  In this case, the cailbration ratio $ r^{(c)}_{\lambda_{s_e,g_2}, \lambda_{n,g_1} }$ need in principal only be measured once and at any temperature. 
In practice, the calibration ratio is weakly temperature-dependent, and a range of temperatures may be specified where calibration will achieve the desired accuracy.  If it is not possible or practical to choose a state $\ket{c}$ which meets the self-calibration criterion, it is still possible to drive a transition from $\ket{g_0}$ to $\ket{c} \neq \ket{s_g}$  to perform a self-consistent calibration at one or more known temperatures.

\begin{figure*}
    \centering
    \includegraphics[width=\textwidth]{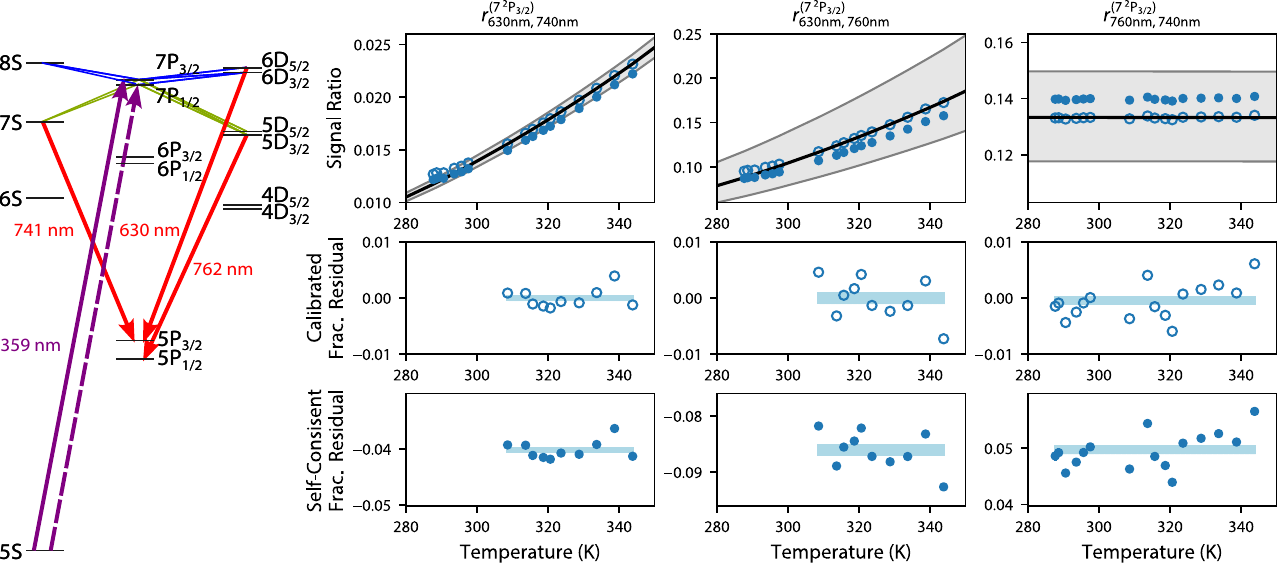}
    \caption{Fluorescence intensity ratio thermometry scheme for Rb.  (Left) Laser light drives the $5^2\rm{S}_{1/2}\rightarrow 7^2\rm{P}_{3/2}$ transition (solid purple arrow).
    Further excitation, primarily due to BBR is shown in blue, while primarily spontaneous decays are shown in green.
    Red arrows show the fluorescent decays which are monitored in this work.
    Self-consistent calibration is performed by tuning the laser to the $5^2\rm{S}_{1/2}\rightarrow 7^2\rm{P}_{1/2}$ transition (dashed purple arrow).
    (Right) Fluorescence ratios (top), calibrated fractional residuals (middle), and self-consistent calibrated fractional residuals (bottom) when driving the  $5^2\rm{S}_{1/2}\rightarrow 7^2\rm{P}_{3/2}$ transition.
    Black lines and gray bands depict the theoretically-predicted ratios and their $1\sigma$ uncertainty, respectively.  
    Open circles show the data calibrated to the predicted ratio. 
    Solid circles show the same data using the self-consistent calibration. 
    Statistical error bars are much smaller than the data points in all panels.
    Light blue bands show the mean value and standard error of the fractional residuals over the highlighted temperature range.
    Vertical scale of the fractional residuals is chosen to highlight the lack of structure above 300~K.
    Fractional residuals for temperatures below 300~K are off-scale and more positive for $r^{(7^2\rm{P}_{3/2})}_{630\,\rm{nm}, 740\,\rm{nm}}$ and $r^{(7^2\rm{P}_{3/2})}_{630\,\rm{nm}, 760\,\rm{nm}}$.}
    \label{fig:Rb}
\end{figure*}



The relevant states for laser excitation, BBR-sensing, and fluorescence detection of Rb used in this work are summarized in Fig.\,\ref{fig:Rb}.   The laser is tuned to fluorescence peak of the $^{85}$Rb  $\ket{g_0} = \ket{5^2\text{S}_{1/2}, F = 3} \rightarrow \ket{s_g} = \ket{7^2\text{P}_{3/2}}$ transitions.  
(The spectroscopic notation $n^2L_J$ specifies the principal quantum number $n$, oribtal angular momentum $L$, and total angular momentum $J$.  $F$ is the total angular momentum.).
In order to detect BBR resonant with the 12.2\,$\mu$m transitions $\ket{s_g} = \ket{7^2\text{P}_{3/2}} \rightarrow \ket{s_e} = \ket{6^2\text{D}_{3/2, 5/2}}$, an interference filter with center wavelength 630\,nm selects fluorescence from the temperature-dependent transition $\ket{s_e} = \ket{6^2\text{D}_{3/2, 5/2}} \rightarrow \ket{g_2} = \ket{5^2\text{P}_{3/2}}$.
Two temperature-independent transitions are monitored.  The first is monitored using an interference filter with center wavelength 760\,nm  to select  $\ket{n_1} = \ket{5^2\text{D}_{3/2, 5/2}} \rightarrow \ket{g_1} = \ket{5^2\text{P}_{1/2}}$ fluorescence.  The second is monitored using  an interference filter with center wavelength 740\,nm  to select $\ket{n_2} = \ket{7^2\text{S}_{1/2}} \rightarrow \ket{g_3} \equiv \ket{g_2}= \ket{5^2\text{P}_{3/2}}$ fluorescence.

We model the  Rb atomic system as a function of radiometric temperature by solving a system of rate equations to determine the steady-state values of $p_i^{(k)}$.  The rate equation model includes spontaneous decay rates $\Gamma_{ij}$ and BBR-stimulated transition rates $\Omega_{ij}^{\rm{BBR}}$ between states $\ket{i}$ and $\ket{j}$, as well as a single laser excitation rate $\Omega_{g_0, k}$.  To the accuracy of this study, the ratio $ r^{(k)}_{\lambda_{s_e,g_2}, \lambda_{n,g_1} }$ is found to not depend on the value of $\Omega_{g_0, k}$, including whether the $\ket{g_0}\rightarrow \ket{k}$ transition is in the linear or saturated regime.  Transition dipole matrix elements are assigned using a modified version of the Alkali Rydberg Calculator python package \cite{Sibalic2017}, with TDME values for all relevant states taken from the Portal for High-Precision Atomic Data and Computation  \cite{UDportal}.
For optically excited states $\ket{k} = \{ 7^2\rm{P}_{1/2}, 7^2\rm{P}_{3/2} $\}, we include all states with $n \leq 10$, $L\leq 3$, and $J=|L\pm1/2|$.

Our CoBRAS fluorescence detection scheme intentionally has insufficient resolution to distinguish hyperfine structure, and the atomic system may be accurately modeled while mostly ignoring the role of nuclear spin $I$.  However, it is important to account for the effect of nuclear spin statistics on the laser excitation.  Laser excitation does not populate all hyperfine Zeeman sublevels. Assuming laser excitation is the predominant mechanism for transfer into the optically excited state ($\ket{s_g}$ or $\ket{c}$), this state is modeled as $n_b$ degenerate bright sublevels which are coupled to the laser, and and $n_d$ degenerate dark sublevels which are not.
Specifically for  linear polarized light resonant with the $^{85}$Rb $\ket{5^2\text{S}_{1/2}, F} \rightarrow \ket{7^2\text{P}_{J^\prime}}$ transition with hyperfine levels of total angular momentum $F$ resolved in the ground state, the degeneracy $(2J + 1)(2I+1)$ of $\ket{7^2\text{P}_{J^\prime}}$ should be reduced by a factor of 3/4, 1/2, 11/12, and 3/4 for the case of $(F=2, J'=1/2)$, $(F=2, J'=3/2)$, $(F=3, J'=1/2)$, and $(F=3, J'=3/2)$, respectively.


The experiment is conducted in a polystyrene foam board box.
The enclosed quartz vapor cell is highly emissive for wavelengths $\lambda \gtrsim 2~\mu$m, and the Rb atoms contained within sense the 12.2~$\mu$m BBR emitted from the quartz.
Three industrial platinum resistance thermometers (PRTs) conforming to ASTM E1137 are affixed to the top, middle, and bottom of the box to measure the temperature and its gradient.  
The PRTs were calibrated by the NIST Industrial Thermometer Calibration Laboratory for temperatures between 273.160~K and 343.152~K.  Each PRT has a temperature calibration uncertainty of 3~mK over this range.
A forced-air Peltier element stabilizes the mean temperature reading of the PRTs to within 0.1\,K.
The uncertainty in the vapor cell temperature, as determined by our 3 PRTs, is limited by the measured thermal gradient across the enclosed volume, which is at most 3~K over the temperature range.

Fluorescence was simultaneously monitored on two photon-counting photomultiplier tubes (PMTs).  Identical bandpass interference filters with full widths at half maximum of 10\,nm were used to select fluorescence wavelengths.   
With a fixed  temperature, fluorescence was collected for 34\,s for each possible filter pair.  
PMT counts  were recorded with the excitation laser blocked for an equal time interval in order to perform background subtraction.
For higher temperatures where the BBR-induced fluorescence was strongest,  the fluorescence ratio shot noise was as low as $\delta r/r = 0.3$\,\% for one second of measurement time.  For all fluorescence ratios, shot noise was less than 0.1\,\%  after a single cycle of the filters.  The observed fluorescence ratios reported here are the geometric mean of the two PMT count rates \cite{Fleming1986}; analyzing each PMT individually yielded results which were consistent with the geometric mean but with roughly 50\,\% higher standard deviations due to reduced counting statistics.

We present two analysis strategies for the CoBRAS.
In the first, we infer the ratio of total detection efficiencies $\eta_{\lambda_{s_e,g_2}}/ \eta_{\lambda_{n,g_1} }$ from a least-squares fit using the observed $r^{(s_g)}_{\lambda_{s_e,g_2}, \lambda_{n,g_1} }$ and the temperature.    In Fig.\,\ref{fig:Rb} these calibrated data are depicted by the open circles.  The rate equation model is depicted by the black line.
The gray band shows the model's $1\sigma$ uncertainty derived from the literature TDME uncertainties \cite{UDportal}.

The calibration is restricted to temperatures between $308$\,K and $344$\,K where the agreement of the  observed fluorescence ratios with our model is excellent.
The root mean square (rms) of the residual is 0.16\,\% of  the predicted ratio $r^{(7^2\rm{P}_{3/2})}_{630\,\rm{nm}, 740\,\rm{nm}}$.  
Assuming only statistical uncertainty results in a reduced $\chi^2$ statistic of 6, indicative of additional noise a few times larger than shot noise. 
The performance is slightly worse for the ratio $r^{(7^2\rm{P}_{3/2})}_{630\,\rm{nm}, 760\,\rm{nm}}$, where the observed rms of the residual is 0.3\,\% of the predicted ratio, and the reduced $\chi^2$ statistic is 22. 
The main result of this investigation is that over the quotidian temperature range $308$\,K to $344$\,K, we realize a calibrated thermometer with a precision of  $\delta T = \delta r \frac{\partial T}{\partial r} \approx 0.13$~K (or fractionally, $\delta T/T \approx 0.04\,\%)$ by measuring the fluorescence ratio 
$r^{(7^2\rm{P}_{3/2})}_{630\,\rm{nm}, 740\,\rm{nm}}$, consistent with the 0.1~K temperature stability of the vapor cell's enclosure.   The temperature accuracy of 3~K is limited by the observed temperature gradient of the temperature controlled enclosures.
By measuring $r^{(7^2\rm{P}_{3/2})}_{630\,\rm{nm}, 760\,\rm{nm}}$, we realize $\delta T = 0.28$~K precision and the same 3~K accuracy. 

The agreement of the measurement with the model is worse when the enclosure is cooled below room temperature.
The observed signal-to-normalization fluorescence ratios systematically increase relative to the model at lower temperatures.
The deviation of the  calibrated data from the model is as much as 8~\% for both $r^{(7^2\rm{P}_{3/2})}_{630\,\rm{nm}, 740\,\rm{nm}}$ and  $r^{(7^2\rm{P}_{3/2})}_{630\,\rm{nm}, 760\,\rm{nm}}$ at $T=286$~K. 
This deviation may be due to leakage room light into the temperature-controlled enclosure, possible condensation of water vapor onto the vapor cell, and systematic differences in the operation of the forced-air Peltier element between heating and cooling operations. 
These effects could be largely mitigated with an improved thermal environment, such as a dry well or air bath.

The ratio $r^{(7^2\rm{P}_{3/2})}_{760\,\rm{nm}, 740\,\rm{nm}}$ provides a consistency check.
As a ratio of two normalization wavelengths, it should be  essentially temperature-independent, and indeed  it is over our entire realized temperature range 286~K to 344~K with an rms deviation of 0.3\,\% of  the mean.  The rate equation model predicts a 0.06\,\% decrease of $r^{(7^2\rm{P}_{3/2})}_{760\,\rm{nm}, 740\,\rm{nm}}$ over this temperature range. 
Thus, we do not observe a  temperature-dependent systematic below room temperature in the state normalization similar to that observed for the $r^{(7^2\rm{P}_{3/2})}_{630\,\rm{nm}, 740\,\rm{nm}}$ and $r^{(7^2\rm{P}_{3/2})}_{630\,\rm{nm}, 760\,\rm{nm}}$ ratios .

We now turn to a second analysis of the fluorescence intensity ratios which provides a self-consistent check of the rate equation model's accuracy and demonstrates some of the aspects of the primary thermometry technique proposed here.
To realize a primary thermometer, the ideal temperature-independent self-calibration states for our scheme are $\ket{c} = \ket{8^2\text{P}_{1/2}}$ or $\ket{8^2\text{P}_{3/2}}$, as these are the next-higher energy states with dipole allowed transitions to the relevant fluorescence states.  
Unfortunately, our frequency doubled Ti:sapphire laser system cannot produce the transition wavelengths from the ground state, around 335\,nm. 
Instead, we tune the laser to the fluorescence peak of the $^{85}$Rb  $\ket{g_0} = \ket{5^2\text{S}_{1/2}, F = 3} \rightarrow \ket{c} = \ket{7^2\text{P}_{1/2}}$ transitions.  
Using these additional data, we infer the ratio of total detection efficiencies $\eta_{\lambda_{s_e,g_2}}/ \eta_{\lambda_{n,g_1} }$ using Eq.\,\eqref{eq:ratio of ratios}, the measured $r^{(c)}_{\lambda_{s_e,g_2}, \lambda_{n,g_1}}$, and the measured temperature.
We refer to this as a ``self-consistent'' calibration of the thermometer because it follows the same self-calibration procedure required for primary thermometry except the choice of calibration state  $\ket{c}$ requires knowledge of $T$.  While the self-consistent calibration is not a primary thermometry technique, its accuracy is a test of our rate equation model and is indicative of the accuracy which might be expected for a primary measurement, {\it e.g.} when using $\ket{c} = 8^2\rm{P}_{3/2}$ in Rb.

Self-consistent calibration using $\ket{c} = \ket{7^2\text{P}_{1/2}}$ (shown as filled circles in Fig.\,\ref{fig:Rb}) results in an overall underestimation of the observed $r^{(7^2\rm{P}_{3/2})}_{630\,\rm{nm}, 740\,\rm{nm}}$ ratio by approximately 4\,\%, and of $r^{(7^2\rm{P}_{3/2})}_{630\,\rm{nm}, 760\,\rm{nm}}$ by approximately 9\,\%.  
This leads to a systematic offset for  the $r^{(7^2\rm{P}_{3/2})}_{630\,\rm{nm}, 740\,\rm{nm}}$ ratio of $\Delta T =-3.5$~K, corresponding to temperature accuracy 1.1\,\%.  For the ratio $r^{(7^2\rm{P}_{3/2})}_{630\,\rm{nm}, 760\,\rm{nm}}$, $\Delta T =  -7.5$~K, corresponding to an accuracy of 2.4\,\%.  
The discrepancy between the self-consistent calibration and the model exceeds the temperature uncertainty of the temperature-controlled enclosure, but is within the model uncertainty for  both  ratios, determined by the TDME uncertainties \cite{UDportal}.  For  $r^{(7^2\rm{P}_{3/2})}_{630\,\rm{nm}, 740\,\rm{nm}}$, all relevant TDMEs have theoretical uncertainties of roughly 0.2\,\% and combine for a total temperature uncertainty of 1.1\,\%.  For $r^{(7^2\rm{P}_{3/2})}_{630\,\rm{nm}, 760\,\rm{nm}}$, 
Ref.\,\cite{Safronova2004} points out the higher theoretical TDME uncertainty for $np - n^\prime d$ transitions versus $np - n^\prime s$ transitions is due to larger electron correlation contributions.
The combined uncertainty contribution of the theoretical TDMEs to the ratio $r^{(7^2\rm{P}_{3/2})}_{630\,\rm{nm}, 760\,\rm{nm}}$ is 8.4\,\%.
The experimental uncertainty is worse than the theoretical uncertainty for the value of the radiative lifetime of the $6^2\text{D}_{3/2}$ state, with roughly 20\,\% disagreement between Ref.\,\cite{Ekers2000} and Ref.\,\cite{vanWijngaarden1992}; $6^2\text{D}_{3/2}$ along with  $6^2\text{D}_{5/2}$, represent $\ket{s_e}$ in our sensing scheme using Rb.  We note that the largely temperature-independent ratio ratio $r^{(7^2\rm{P}_{3/2})}_{760\,\rm{nm}, 740\,\rm{nm}}$ exceeds the modeled value by roughly 5\,\%, but is well within the theoretical uncertainty when using the self-consistent calibration.

Table \ref{tab:error budget} summarizes the error budget for the self-consistent calibrated CoBRAS.
Uncertainty in the values of TDMEs is a leading uncertainty for the temperature measurement.
Temperature uncertainty is estimated as the maximum measured thermal gradient of 3~K, which could be made negligible by shrinking the volume of the thermal enclosure.
Depending on the atomic temperature, density, and transition involved, state-changing collisions may either increase or decrease the observed signal-to-normalization fluorescence ratio. However, we observe no structure in the residuals of the observed fluorescence ratios when comparing to the rate equation model, and therefore conclude that the net effect of collisions is less than the precision of the current measurement for temperatures up to 344~K.  The effect of collisions on the temperature measurement are discussed in further detail in the Supplemental Material.  
To keep the pulse pileup correction below a 1\,\%, the laser power is adjusted so that the maximum photon count rate on any detection channel is less than $10^6$\,$\text{s}^{-1}$, approximately 100 $\mu$W.  
Uncertainty due to this correction is $<0.1\,$\%.
Self-heating effects are negligible as $<1$~$\mu$W of laser power is absorbed.
Radiation trapping can also distort the BBR measurement. 
In particular, monitoring transitions to the ground state $\ket{g_0}$ or other low-lying states with substantial thermal population should be avoided. 
Such emitted photons must traverse the entire atomic vapor, thus making the optical depth particularly high.  
On the other hand, the signal and normalization decay photons will only experience radiation trapping within the volume of the laser beam enlarged by the characteristic distances $\Tilde{v}/ \Gamma_{g_2}$ and $\Tilde{v}/\Gamma_{g_1}$, where $\Tilde{v}$ is the mean thermal velocity of the atoms. 
This additional optical path length can be mitigated by choosing $\ket{g_1}$ and $\ket{g_2}$  to have sufficiently fast radiative decay rates. 
For the states depicted in Fig.\,\ref{fig:Rb}, these  characteristic distances are of order 10~$\mu$m, or less than 1\,\% of the laser beam diameter.  
Thus, appreciable radiation trapping occurs only within the volume of the exciting laser beam.
We estimate the effect of radiation trapping to be less than 0.2~\%.



We have demonstrated a Compact Blackbody Radiation Atomic Sensor (CoBRAS) based on Rb with several desirable characteristics for a practical device.  The device realizes a temperature precision of $\delta T = 0.13$~K from 308~K to 344~K, or a fractional precision of $\delta T/T = 0.04$\,\%.  The temperature measurement is fast, with statistical uncertainty as low as  0.1~\% in one second. The device is described by a simple rate equation model.  The equipment required is also quite modest: a single laser and a few commercial stock components (alkali vapor cell, bandpass filters, and PMTs).  The accuracy of the calibrated device is entirely limited by the technical limitations of the thermal enclosure, and substantial improvements should be straightforward.

\begin{table}
\caption{Self-consistent calibrated CoBRAS $1\sigma$ error budget.}

\begin{tabular}{l cc}

\hline\hline
            & \multicolumn{2}{c}{$u(T)/T$ (\%)}  \\
Uncertainty                         &$r^{(7^2\rm{P}_{3/2})}_{630\,\rm{nm}, 740\,\rm{nm}}$  & $r^{(7^2\rm{P}_{3/2})}_{630\,\rm{nm}, 760\,\rm{nm}}$ \\
\hline
TDME \cite{UDportal}             & 1.1 & 8.4                     \\

Temperature Gradient & \multicolumn{2}{c}{\quad1.0}    \\

Radiation Trapping      &\multicolumn{2}{c}{\quad0.2}  \\
Collisions              & \multicolumn{2}{c}{$< 0.1$}                                      \\
Pulse Pileup            &\multicolumn{2}{c}{ $< 0.1$}                                      \\  
Statistical            &\multicolumn{2}{c}{ $< 0.1$}       
\vspace{2pt}\\       
Total (quadrature sum)                   & 1.5 &    8.5                         \\
\hline \hline
\end{tabular}

\label{tab:error budget}
\end{table}

We also demonstrate a self-consistent calibrated thermometer as a step towards realizing a primary thermometer. We observe systematic offsets from the reference thermometer readings of a few kelvin, consistent with the reported theoretical uncertainty in the TDME's of the model \cite{UDportal}. Future work will investigate systematic effects in a precision temperature-controlled environments, such as a dry well.  Using such a high-accuracy temperature reference (typically 10\,mK), it should be possible to improve experimental TDME uncertainties, such that a self-calibrated CoBRAS approaches or exceeds the 0.04\,\% temperature precision demonstrated here.

Finally, it should be noted that a CoBRAS may be designed to operate as either a contact or non-contact thermometer for a desired application.
Our quartz vapor cell is highly emissive for wavelengths $\lambda \gtrsim 2~\mu$m.  
Thus the implementation described here is a contact thermometer, where atoms sense 12.2~$\mu$m BBR emitted from the vapor cell, which comes to thermal equilibrium with its environment.
Atomic vapor cells with volumes less than 1~mm$^3$ are commercially available for chip-scale sensors, which could enable fit-for-purpose \mbox{CoBRAS} contact thermometer replacements for platinum resistance thermometers.
Conversely, one can engineer a non-contact thermometer where atoms sense the spatially averaged radiometric temperature of the environment by changing the cell material, ground sensing state, or excited sensing state such that the cell is highly transmissive at the sensed BBR wavelength.
Sapphire \cite{Jau2020}, CaF$_2$, ZnSe, and chemical vapor deposition diamond are some possible choices for vapor cell construction for non-contact fluorescence thermometry with alkali atoms.

\section*{Acknowledgments}
The authors thank Nickolas Pilgram, Daniel Barker, and Joe Rice for useful discussions and Zeeshan Ahmed, Dixith Manchaiah, and Wes Tew for a thorough reading of the manuscript.  DSL was supported by a National Research Council Postdoctoral Fellowship.  This work was supported by NIST, and by DARPA under the SAVaNT program.  The views, opinions and/or findings expressed are those of the authors and should not be interpreted as representing the official views or policies of the Department of Defense or the U.S. Government. A contribution of the U.S. government, this work is not subject to copyright in the United States.

\subsection*{Conflict of Interest}
\vspace{-3mm}
The authors have no conflicts to disclose.
\vspace{-3mm}
\subsection*{Data Availability Statement}
\vspace{-3mm}
All data presented here will be made available to the reader upon reasonable request to the authors.

\bibliography{main}

\appendix\label{sec:SM}
\section*{Supplemental Material}


\begin{figure}
    \centering
    \includegraphics[width=\linewidth]{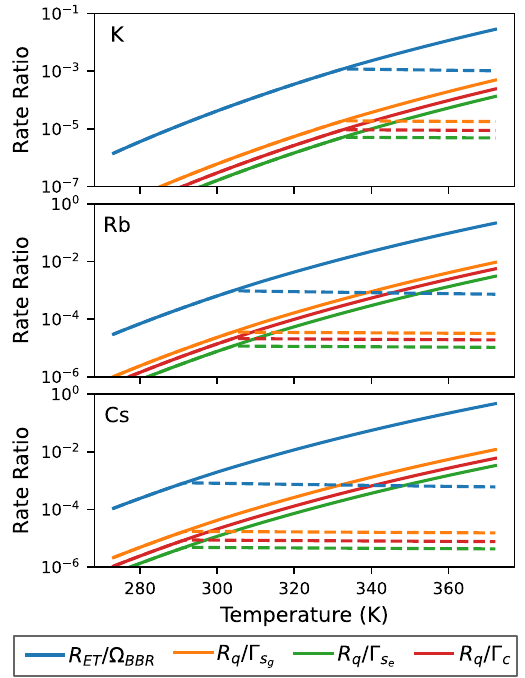}
    \caption{Ratios of state-changing collision rates to BBR-stimulated and spontaneous rates for K, Rb, and Cs atoms as a function of temperature.
    Taking $n_{\rm{g}}$ as the principal quantum number of the ground state for each atom, we model a thermometry scheme where $\ket{g_0} = n_{\rm{g}}\rm{S}_{1/2}$,  $\ket{s_g} = (n_{\rm{g}}+2)\rm{P}_{3/2}$, $\ket{s_e} = (n_{\rm{g}}+3)\rm{S}_{1/2}$, $\ket{n} = (n_{\rm{g}}+2)\rm{S}_{1/2}$, and $\ket{g_1} =\ket{g_2} = n_{\rm{g}}\rm{P}_{3/2}$.
    The relevant sensed BBR wavelengths are  7.9~$\mu$m, 8.5~$\mu$m, and 8.9~$\mu$m, respectively. 
    For the model, we assume all state-changing collision cross-sections are $\sigma_{\rm{scc}} = 10^{-13}$\,cm$^2$.
    The solid curves assume the density given by the saturated vapor pressure \cite{Alcock1984,Sibalic2017}; dashed curves assume a starved cell with maximum density of $2.5\times 10^{10}$~cm$^{-3}$.
    }
    \label{fig:collisions}
\end{figure}

To avoid additional population dynamics beyond the BBR-stimulated transitions and spontaneous decay included in our model, state-changing collisions must be minimized.  
The rate of state-changing collisions for a given excited state is $R_{\rm{scc}} = n \langle\sigma_{\rm{scc}} v \rangle$, where $n$ is the atomic number density, $\sigma_{\rm{scc}}$ is the total  state-changing collision cross-section of the state, $v$ is the center of mass velocity, and $\langle\dots\rangle$ denotes a thermal average. 
The total  state-changing collision cross-section comprises  excitation transfer, $J$-changing,  and quenching processes \cite{Pace1974,Keramati1988,Ekers2000,Bieniak2000,Glodz2019}. For the first few excited states of an alkali atom, each of these processes may have a  cross-section as large as $\sim 10^{-13}$\,cm$^{2}$ \cite{Ekers2000}.

Depending on the atomic temperature, density, and transition involved, state-changing collisions may either increase or decrease the observed signal-to-normalization fluorescence ratio. 
Figure \ref{fig:collisions} illustrates the relative importance of the excitation transfer, $J$-changing,  and quenching processes for thermometry schemes in K, Rb, and Cs.  A saturated vapor pressure and a hypothetical $\sigma_{\rm{scc}} =10^{-13}$\,cm$^{2}$ is assumed for all state-changing collisions.  Taking $n_{\rm{g}} = 4,5,6$  as the principal quantum number of the ground state for each respective atom, we model a thermometry scheme where $\ket{g_0} = n_{\rm{g}}\,^2\rm{S}_{1/2}$,  $\ket{s_g} = (n_{\rm{g}}+2)^2\rm{P}_{3/2}$, $\ket{s_e} = (n_{\rm{g}}+3)^2\rm{S}_{1/2}$, $\ket{n} = (n_{\rm{g}}+2)^2\rm{S}_{1/2}$, $\ket{g_1} = n_{\rm{g}}\,^2\rm{P}_{3/2} $ and $\ket{g_2} = n_{\rm{g}}\,^2\rm{P}_{1/2}$. 
For the example thermometry schemes in Fig.\,\ref{fig:collisions}, the BBR wavelengths are 7.9~$\mu$m, 8.5~$\mu$m, and 8.9~$\mu$m, for K, Rb, and Cs, respectively.

Excitation transfer collisions can mimic the BBR-simulated $\ket{s_g}\rightarrow \ket{s_e}$ transition, and thus increase the signal-to-normalization fluorescence ratio.
A colliding pair of atoms must have a center-of-mass kinetic energy which exceeds the $\ket{s_g}\rightarrow \ket{s_e}$ transition energy in order for excitation transfer to occur.  
This energy threshold suppresses the the excitation transfer rate $R_{\rm{ET}}$ compared to other state-changing collision rates.
For the experimentally studied $T<344$~K of this work and the modeled assumptions, the ratio $R_{\rm{ET}}/\Omega_{\rm{BBR}}$ is less than 0.3~\% for K, 3~\% for Rb, and 7~\% for Cs.  

Quenching collisions will reduce fluorescence from all excited states.  
The observed signal-to-normalization fluorescence ratio can thus be increased by quenching collisions in $\ket{n}$, decreased by  quenching collisions in $\ket{s_e}$, or decreased by de-excitation $\ket{s_g}\rightarrow \ket{n}$.
Therefore, the net effect of quenching collisions on the fluorescence thermometry measurement will depend on the relative magnitudes of all quenching collision cross-sections affecting the excited states.
The magnitude of the quenching collision rate $R_{\rm{q}}$  should be compared to the spontaneous decay rate $\Gamma_i$ of each excited state $i$ which emits a monitored fluorescence wavelength.
For the experimentally studied $T<344$~K of this work and the modeled assumptions, the ratio $R_{\rm{q}}/\Gamma$ is less than 0.02~\% for K, 0.2~\% for Rb, and 0.4~\% for Cs.  

We note that the temperature dependence of state-changing collision systematic effects is dominated by the rapid change in saturated atomic density.
The dashed lines in Fig.\,\ref{fig:collisions} show the ratios of state-changing collision rates to BBR-stimulated and spontaneous rates using a starved vapor cell with a maximum atomic density of $2.5\times 10^{10}$~cm$^{-3}$ ({\it i.e.}, the saturated vapor pressure of Cs at $T=293$~K).  
In this case all collisional effects are at most 0.1\,\%.

The temperature range explored here was limited entirely by technical considerations in setting the temperature.  Surprisingly, no deviation from the modeled e fluorescence ratios was observed even at our highest attainable temperature where the above estimates  predict state-changing collisions may be detectable.  An important next step will be to measure fluorescence ratios at higher temperatures to observe where collisions become relevant.  Collisional effects could be added to the rate equation model in a straightforward manner.  

\end{document}